\documentclass[%
 aip,
 amsmath,amssymb,
 reprint,%
]{revtex4-1}

\usepackage{graphicx}% Include figure files
\usepackage{dcolumn}% Align table columns on decimal point
\usepackage{bm}% bold math
\usepackage[utf8]{inputenc}
\usepackage[T1]{fontenc}
\usepackage{mathptmx}
\usepackage{etoolbox}

\usepackage{easyReview}
\usepackage{todonotes}
\usepackage{subcaption}
\usepackage{chemformula}
\usepackage{textcomp}

\graphicspath{{./img/}}

\makeatletter
\def\@email#1#2{%
 \endgroup
 \patchcmd{\titleblock@produce}
  {\frontmatter@RRAPformat}
  {\frontmatter@RRAPformat{\produce@RRAP{*#1\href{mailto:#2}{#2}}}\frontmatter@RRAPformat}
  {}{}
}%
\makeatother
\begin{document}

\preprint{AIP/123-QED}

\title{Continuum model for extraction and retention in porous media}
% Force line breaks with \\
\author{Andr\'{e} F.~V. Matias}
 \email{Corresponding author afmatias@fc.ul.pt}
 \affiliation{Centro de F\'{i}sica Te\'{o}rica e Computacional, Faculdade de Ci\^{e}ncias, Universidade de Lisboa, 1749--016 Lisboa, Portugal}
 \affiliation{Departamento de F\'{i}sica, Faculdade de Ci\^{e}ncias, Universidade de Lisboa, 1749--016 Lisboa, Portugal}
 
\author{Daniel F. Valente-Matias}
 \affiliation{Centro de Qu\'{i}mica Estrutural, Institute of Molecular Sciences, Faculdade de Ci\^{e}ncias, Universidade de Lisboa, Campo Grande, 1749-016 Lisboa, Portugal}
 \affiliation{Departamento de Qu\'{i}mica e Bioqu\'{i}mica, Faculdade de Ci\^{e}ncias, Universidade de Lisboa, 1749--016 Lisboa, Portugal}
 
\author{Nuno R. Neng}
 \affiliation{Centro de Qu\'{i}mica Estrutural, Institute of Molecular Sciences, Faculdade de Ci\^{e}ncias, Universidade de Lisboa, Campo Grande, 1749-016 Lisboa, Portugal}
 \affiliation{Departamento de Qu\'{i}mica e Bioqu\'{i}mica, Faculdade de Ci\^{e}ncias, Universidade de Lisboa, 1749--016 Lisboa, Portugal}
 
\author{Jos\'{e} M. F. Nogueira}
 \affiliation{Centro de Qu\'{i}mica Estrutural, Institute of Molecular Sciences, Faculdade de Ci\^{e}ncias, Universidade de Lisboa, Campo Grande, 1749-016 Lisboa, Portugal}
 \affiliation{Departamento de Qu\'{i}mica e Bioqu\'{i}mica, Faculdade de Ci\^{e}ncias, Universidade de Lisboa, 1749--016 Lisboa, Portugal}

\author{Jos\'{e} S. Andrade Jr.}
 \affiliation{Departamento de F\'{i}sica, Universidade Federal do Cear\'{a}, 60451--970, Fortaleza, Cear\'{a}, Brazil}

\author{Rodrigo C.~V. Coelho}
\author{Nuno A.~M. Ara\'{u}jo}%
 \affiliation{Centro de F\'{i}sica Te\'{o}rica e Computacional, Faculdade de Ci\^{e}ncias, Universidade de Lisboa, 1749--016 Lisboa, Portugal}%
 \affiliation{Departamento de F\'{i}sica, Faculdade de Ci\^{e}ncias, Universidade de Lisboa, 1749--016 Lisboa, Portugal}

\date{\today}% It is always \today, today,

\begin{abstract}
Several natural and industrial processes involve the extraction or retention of a solute by a fluid invading a network of channels. Examples include aquifer contamination, chemical filtration, and coffee extraction. We propose a continuum equation to model these processes, parametrized by the Péclet number and the rate of mass transfer between the solid and the fluid. We study the time dependence of the extracted mass for different values of the parameter space. The continuum description is validated by combining extraction experiments with coffee and computational fluid dynamics. An analytical solution is derived for the limit of slow mass transfer, which is corroborated by numerical simulations.
\end{abstract}

\maketitle

\section{Introduction}

Coffee is among the most popular beverages around the world, with a worldwide consumption of more than one million kilograms per hour~\cite{StatistaCoffee2021}. Over five centuries, the chase for the best coffee has relied mainly on empirical investigation. However, more recently, there has been an effort towards a more systematic approach, considering the fluid flow~\cite{Sano2019, Giacomini2020}, dissolution~\cite{Kuhn2017,Espinoza-Perez2007}, mass transport~\cite{Corrochano2015, Ellero2019}, and changes in the porous medium through swelling~\cite{Mo2022} and erosion~\cite{Mo2021}, with a remarkable impact in the coffee industry~\cite{Cameron2020, QYResearch2019}.

In coffee extraction water percolates through the interstices of a pack of milled coffee beans, dissolving a blend of solutes, which give to the coffee drink the colour, aroma, and taste~\cite{Illy2005, Bear2018}. As the fluid flows, the porous structure changes in an interplay between grain swelling, dissolution, and erosion~\cite{Matias2021, Mateus2007, Corrochano2015}. The quality of the extracted coffee depends on the tuning of variables such as the water flow rate and temperature, and the level of compaction and milling of the beans. Thus, an accurate prediction of the solute concentration in the cup is a non-trivial endeavor, for the solute dissolution and the consequent transport depends on the entire dynamics.

Reactive porous medium is also a relevant topic for other applications, as in the development of catalysts~\cite{Chandra2019} and channelization in porous media~\cite{Menke2023}. Several works propose continuum models for reactive porous media using upscaling methods~\cite{Ladd2021}. But they mostly focus on regular lattices that are more easily handled analytically~\cite{Guo2015, Mei1992}. For more complex geometries we need to focus on the pore scale. Studies on the transport of dissolved solutes at the pore scale~\cite{Ellero2019, Mo2022} provide valuable information about the mechanics of dissolution and subsequent transport across the pores, but they fail to access the larger length and time scales. For that, a coarse grained approach is required. Here, we propose a continuum equation that estimates the amount of caffeine extracted using experimentally measured quantities. Our model is benchmarked with experiments using a capsule espresso machine.

Coffee extraction is an advection-diffusion process, where the competing mechanisms are captured by the Péclet number, a dimensionless number that is high when advection dominates and it is low when diffusion dominates. For an espresso coffee, the extraction occurs at high Péclet ($\mathrm{Pe}\sim 10^5$). Despite focusing on espresso extraction, knowledge gains can be translated into other processes that involve the dissolution of chemical compounds such as subsurface soil contamination~\cite{Abriola1985, Singh2010, Singh2018} and grain drying in silos~\cite{Brooker1992}. For both cases the fluid flow rate is low and thus the Péclet number is lower than in the coffee extraction. Here we propose a continuum model that is valid also for low Péclet. Results obtained with the model agree with pore scale simulations for the solute transport in a regime where diffusion competes with advection ($\mathrm{Pe}\sim 10^2$). For the particular case of grain drying, the Péclet number is low but also the dissolution is very slow, and thus the system is on a quasi-steady state. For this regime, we derive an analytical solution for the proposed mathematical model that agrees with pore scale simulations for both $\mathrm{Pe}\rightarrow 0$ and $\mathrm{Pe}\rightarrow\infty$.

The paper is organized as follows. In Sec.~\ref{sec:model}, we describe the proposed model starting from the generic advection-diffusion equation and modify it to the particular case of dissolution. In Sec.~\ref{sec:results}, we present the obtained results, starting with validation using experimental measurements, followed by the comparison with the pore scale model, and finishing with a comparative study with the regime of very slow dissolution. In Sec.~\ref{sec:conclusion}, we draw some conclusions.

\section{Coffee extraction}
\label{sec:model}

\begin{figure}
	\centering
	\includegraphics[width=\linewidth]{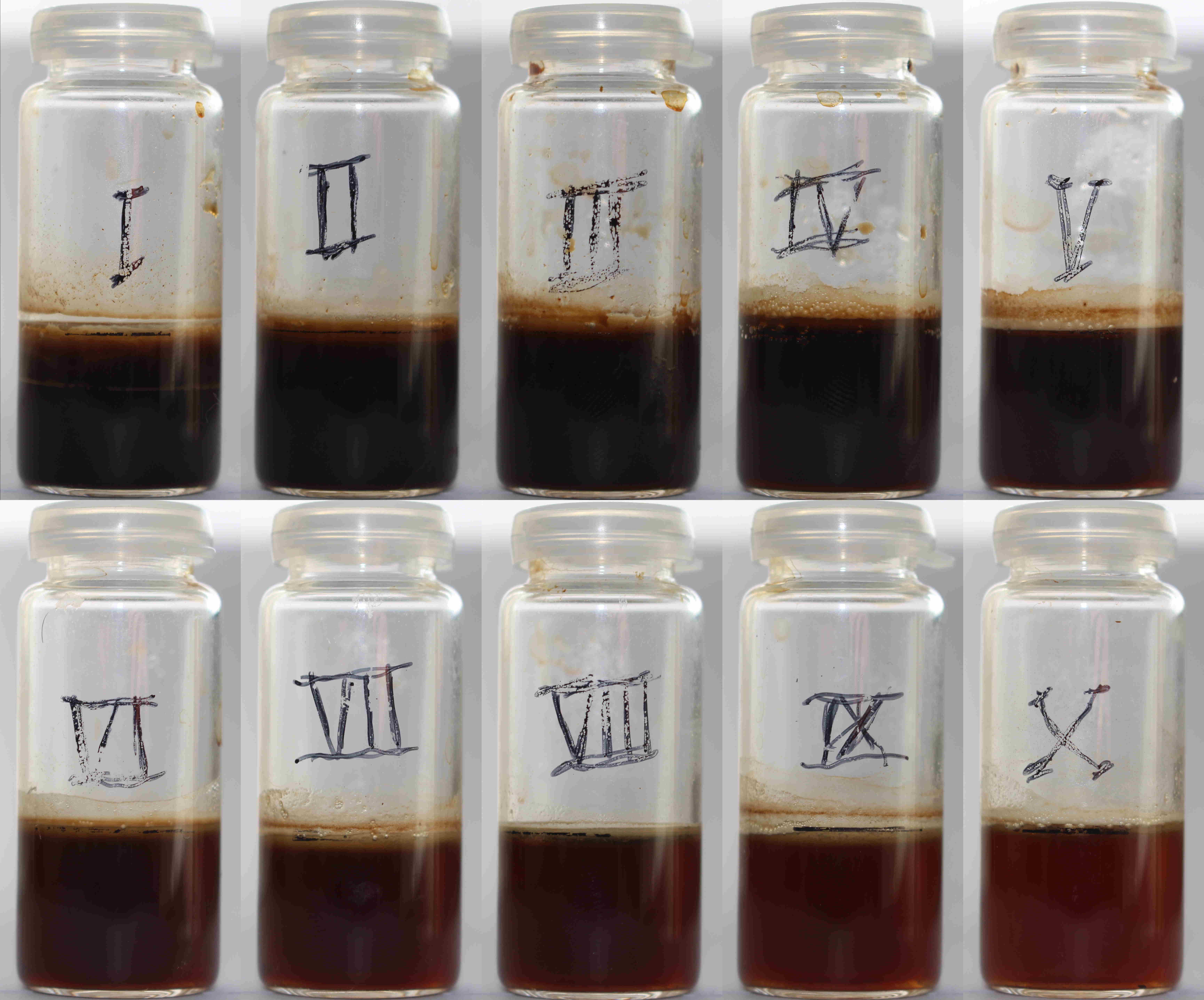}
	\caption{Glass vials with coffee samples extracted from a capsule espresso machine. Each vial contains 5~mL of liquid and it is ordered temporally from I to X.}
	\label{fig:setup}
\end{figure}

The typical coffee espresso is obtained when hot water at 90~\textdegree C and 9~bar is pushed through a pack of 6.5~g of ground coffee. The process takes between 25 and 30 seconds resulting in 25-30~mL of beverage~\cite{Illy2005}. Using a capsule espresso machine, we can consistently reproduce the process and measure the desired quantities. Here we focus on caffeine, given the large number of solutes present in coffee. We measured the time evolution of the caffeine transport through the coffee by extracting a long espresso and taking samples every 5~mL, as seen in Fig.~\ref{fig:setup}. The caffeine content was measured using High Performance Liquid Chromatography (HPLC). To measure the total caffeine content of the ground beans, a full extraction was also performed using Soxhlet extraction. The details of the experimental protocol can be found in the Methods appendix (Appendix~\ref{app:experiment}).

The colour of the liquid in Fig.~\ref{fig:setup} is lighter for later samples, suggesting that the solute concentration decreases with the volume extracted. Later we show that this qualitative observation is corroborated by data. Thus, increasing the beverage volume increases the caffeine content at the expense of concentration, usually referred to as strength. Next, we propose a continuum equation, based on the advection-diffusion equation, to model the caffeine extraction and compare it to the experimental data.

\subsection{Advection-diffusion equation}

The transport of a dissolved solute through a porous medium is driven by fluid drag (advection) and Brownian motion (diffusion). This, together with dissolution, is combined in the advection-diffusion equation:
\begin{equation}
	\dot{C}(\mathbf{x}, t) + \nabla \cdot \left[\mathbf{u}(\mathbf{x}, t) C(\mathbf{x}, t)\right] = \nabla \cdot \left[D(\mathbf{x}) \nabla C(\mathbf{x}, t)\right] + S(\mathbf{x}, t) \textrm{ ,}
	\label{eq:ade}
\end{equation}
where on the left-hand side, we have the time evolution of the concentration and the advection term and on the right-hand side the diffusion term and the reaction term, which accounts for the dissolution in our case. Furthermore, $\mathbf{x}$ is the position, $t$ is time, $C$ is the solute concentration, $\mathbf{u}$ is the interstitial fluid velocity, $D$ is the diffusion coefficient, and $S$ is the rate of dissolution. We consider the spatial average, across the representative element volume (REV), fluid velocity $\mathbf{u}_m(t) = \lVert \mathbf{u} \rVert \hat{e}_v$, where $\lVert \mathbf{u} \rVert$ and $\hat{e}_v$ are the velocity magnitude and direction respectively, thus considering the microscopic variations of the fluid velocity in the hydrodynamic dispersion~\cite{Bear2018}. The diffusion coefficient, in general a tensor, here is assumed to be isotropic and uniform. Note that, there is dispersion due to hydrodynamic effects. For a solute transport between parallel plates separated by a distance $L$, the coefficient of hydrodynamic dispersion can be calculated as~\cite{Taylor1953, Aris1956, Valdes-Parada2011}
\begin{equation}
	D^* = D \left[1 + \left(\frac{\lVert \mathbf{u} \rVert L}{D}\right)^2 \frac{1}{210}\right] \textrm{ .}
	\label{eq:Deff_parallel}
\end{equation}
Thus, for $\lVert \mathbf{u} \rVert = 0 $ the hydrodynamic dispersion is $ D^* = D $, while the advection mechanism dominates for high values of the velocity and $ D^* \sim \lVert \mathbf{u} \rVert^2 $.

We also consider that dissolution only occurs when the grains are surrounded by water. Hence, the pores which start filled with air are invaded by the water enabling dissolution. For simplicity, we assume that the waterfront is a straight line always perpendicular to the average fluid velocity (plug-like), as represented with the dashed lines in Fig.~\ref{fig:interface}(a) for different times, thus the source term $S$ is corrected with the Heaviside function. The fluid invasion could be described in more detail using, for instance, a free surface model at the pore scale\cite{Janssen2010}, but here we only consider a minimal model for its dynamics. Equation~\eqref{eq:ade} simplifies as,
\begin{equation}
	\dot{C} + \mathbf{u}_m \cdot \nabla C = D^* \nabla^2 C + S~\Theta \left(\frac{t}{t^*} - 1 \right) \textrm{ ,}
	\label{eq:ade_hom}
\end{equation}
where $\Theta$ is the Heaviside function that is zero if its argument is negative ($t<t^*$) and one otherwise ($t\ge t^*$), and $t^*$ is the time it takes for the invading fluid to reach the position $\mathbf{x}$ for the first time, \textit{i.e.},
\begin{equation}
	t^* = \frac{\mathbf{x} \cdot \mathbf{u}_m}{\lVert \mathbf{u}_m \rVert^2} \textrm{ .}
	\label{eq:interface_time}
\end{equation}
This simple approach does not limit the solute motion beyond the interface (in the air phase), but it should be valid for cases where advection is the dominant mechanism of mass transport.

\subsection{Dissolution}
\label{sec:theory_diss}

\begin{figure}
	\centering
	\includegraphics[width=\linewidth]{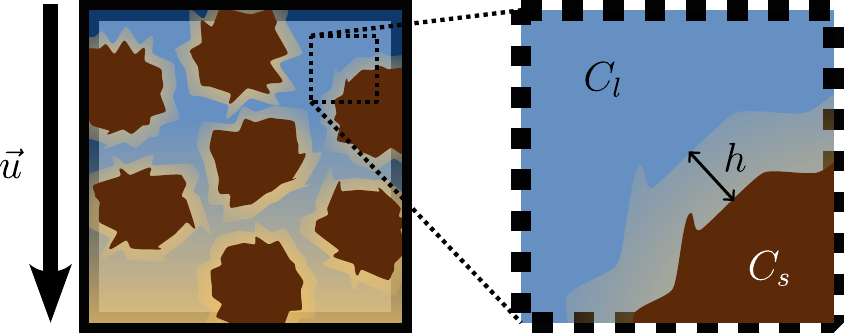}
	\caption{Schematics of the dissolution of grains that compose the porous medium. On the left, the grains are surrounded by liquid, represented in blue and, as they dissolve, the solute concentration in the fluid increases, represented by the shades of yellow. On the right, a detail of a dissolving grain. The concentration gradient across the solid-liquid interface (represented with an infinitesimal width $h$) changes from $C_s$, at the grain surface, to $C_l$, in the liquid.}
	\label{fig:dissolution}
\end{figure}

In caffeine extraction, the source term corresponds to the dissolution, that is caused by the contact between the fluid and the solid matrix, and is driven by diffusion. Thus, the solute flux through the solid boundary is given by~\cite{Tzia2003}
\begin{equation}
	\mathbf{j}_{s\rightarrow l} = - D \nabla C \hat{\mathbf{n}} \textrm{ ,}
\end{equation}
where $\hat{n}$ is the unitary vector perpendicular to the solid surface. In the case of coffee extraction, the dissolved solute is much smaller than the solid matrix, thus we neglect changes in the volume of the structure. Since it is not trivial to deal with spatial derivatives across the interface (the width of the transition region $h$ is non-trivial to measure experimentally, see Fig.~\ref{fig:dissolution}), the common practice is to assume that the mass transferred from the solid to the liquid depends on a mass transfer coefficient, which multiplied by the proper pre-factor gives the rate of mass transfer~\cite{Tzia2003, Kang2003, Rahmat2019}. Here the driving force is diffusion, hence the mass variation $\dot{m}$ at the interface is
\begin{equation}
	\dot{m} = k_c (C_s - C_l) A \textrm{ ,}
	\label{eq:diss}
\end{equation}
where the mass transfer rate coefficient can be estimated as $k_c \equiv D / h $, $C_s$ is the microscopic concentration at the surface, $C_l$ is the concentration in the liquid bulk, and $A$ is the surface area in the REV, as illustrated in Fig.~\ref{fig:dissolution}. The initial value of $ C_s $ determines  mass that can be dissolved
\begin{equation}
	m_T = \int C_s dV \textrm{ ,}
	\label{eq:total_mass}
\end{equation}
where the integral is over the entire medium. The total accessible solute mass is constant thus, we neglect the solute diffusion inside the grains, assumed much slower than the dissolution. This might affect the predictive capabilities for large volumes of extractions. To evaluate the concentration variation at the liquid and solid we need to divide the mass variation by the liquid or solid volume. Thus, the source term in Eq.~\eqref{eq:ade_hom} becomes
\begin{equation}
	S = \frac{\dot{m}}{v_l} = k_c (C_s - C_l) \frac{A}{v_l} \textrm{ ,}
	\label{eq:source_term}
\end{equation}
where $v_l$ is the liquid volume of the REV. To ensure total mass conservation, the solute mass inside the solid needs to change, thus the concentration at the solid-liquid interface evolves as
\begin{equation}
	\dot{C}_s = - \frac{\dot{m}}{v_s} ~\Theta\left(\frac{t}{t^*} - 1 \right) \textrm{ ,}
	\label{eq:Cs_time}
\end{equation}
where $v_s$ is solid volume of the REV where the dissolution occurs. The above equations can also handle buffering by limiting the dissolution when the concentration is close to saturation. Here, we assume that $C_s$ is much smaller than the saturation concentration and thus, buffering is neglected.

Combining Eqs.~\eqref{eq:ade_hom},~\eqref{eq:source_term}, and~\eqref{eq:Cs_time} we can describe the solute transport throughout the liquid and at the solid-liquid interface. These equations can be made dimensionless by applying,
\begin{align}
	\mathbf{x} & \rightarrow L \mathbf{x} \\
	t & \rightarrow \frac{L}{U} t \textrm{ ,}
\end{align}
where $L$ and $U$ are the characteristic length and velocity. Thus, we obtain
\begin{align}
	&\dot{C} + \nabla \cdot C = \frac{1}{\mathrm{Pe}} \nabla^2 C + \xi\frac{\textrm{Sh}}{\mathrm{Pe}}~\Theta \left(\frac{t}{t^*} - 1 \right) \textrm{ ,}
	\label{eq:ade_adim} \\
	&\dot{C}_s = - \xi_s\frac{\textrm{Sh}}{\mathrm{Pe}}~\Theta \left(\frac{t}{t^*} - 1 \right) \textrm{ ,}
\end{align}
where $\xi=A/v_l$ ($\xi_s=A/v_s$) is the ratio between exposed surface area and liquid (solid) volume in the REV. $\mathrm{Pe}$ and $ \textrm{Sh} $ are the Péclet and Sherwood dimensionless numbers, defined as
\begin{equation}
	\mathrm{Pe} = \frac{\textrm{advection}}{\textrm{diffusion}} = \frac{U L}{D} \textrm{ ,}
	\label{eq:peclet}
\end{equation}
and
\begin{equation}
	\textrm{Sh} = \frac{\textrm{mass transfer}}{\textrm{diffusion}} = \frac{k_c L}{D} \textrm{ ,}
	\label{eq:sherwood}
\end{equation}
respectively, where the mass transfer on the Sherwood number corresponds to the transfer from inside of the grains to the liquid. Notice that some literature use the Damköhler instead of the Sherwood number. We found more appropriate to use the later as it deals with mass transfer rates instead of reaction rates.

The initial concentration is zero everywhere, the concentration gradient in the outlet is zero at all times, and inlet liquid, at $\mathbf{x}_0$, has no solute. Since the grains start full of caffeine, the initial solute mass inside the solid can be determined using the total caffeine mass. The initial boundary conditions are then,
\begin{align}
	C(t,\mathbf{x}_0)        & = 0 \textrm{ ,}   
	\label{eq:inlet} \\
	\nabla C(t,\mathbf{x}) \cdot \hat{\mathbf{n}} & = 0 \textrm{ ,}
	\label{eq:outlet} \\
	C(0,\mathbf{x})        & = 0 \textrm{ ,}   \\
	C_s(0,\mathbf{x})      & = C_i \textrm{ ,}
\end{align}
where $\hat{\mathbf{n}}$ is the normal to the outlet. To compare with the experiments, we assume that the solute is evenly spread across the porous medium and so, $C_i$ was determined according to the total caffeine measured experimentally 
\begin{equation}
	C_i = \frac{m_T}{L ~ \pi R_e^2} \textrm{ ,}
\end{equation}
where $m_T$ is the total mass of caffeine measured using the Soxhlet extraction, $L$ is the height of the coffee cake, and $R_e$ is the outlet radius of the espresso machine.

\section{Results}
\label{sec:results}

We first validate the continuum equation with the experimental measurements of caffeine, described in Sec.~\ref{sec:model}, corresponding to a regime of very high Péclet number ($\mathrm{Pe} \sim 10^5$). We follow with the comparison of the continuum equation for low Péclet numbers, where advection dominates but diffusion starts to be relevant, which is done using the lattice Boltzmann method. We finalize with a regime where the transfer rate is so small that the system reaches a steady state, for which we can derive an analytical solutions.

\subsection{Caffeine in an espresso}
\label{sec:coffee}

\begin{figure}
	\centering
	\includegraphics[width=\linewidth]{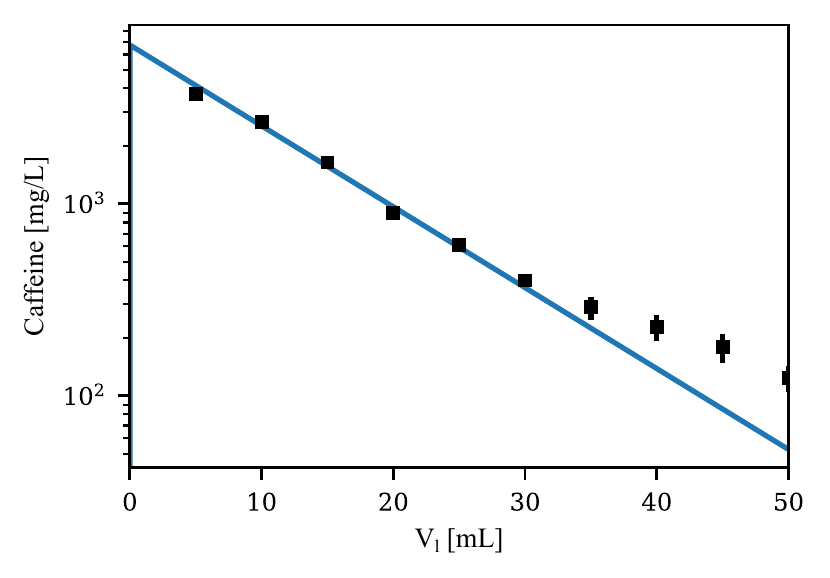}
	\caption{Log-linear plot of the outlet concentration as a function of the volume of liquid extracted. The blue line correspond to the numerical solution of Eq.~\eqref{eq:ade_hom} and the black squares the experimental measurements and the error bars, occasionally smaller than the symbols, to the standard error.}
	\label{fig:experiment}
\end{figure}

The experimentally measured caffeine concentration in the cup corresponds to the ratio of the mass of solute and volume of liquid extracted $m_s/V$. The caffeine concentration in each 5~mL vial as a function of the total liquid extracted $V_l$ is presented in Fig.~\ref{fig:experiment} with black squares. The error bars correspond to the standard error across the four samples. As suggested by the results in Fig.~\ref{fig:experiment}, the caffeine concentration decreases exponentially with $V_l$. For the later stages of extraction, the decay deviates from the exponential, which might stem from a slow down in the extraction at the grain level due to internal diffusion inside the grains.

The parameters for coffee extraction can be measured experimentally and found in the literature~\cite{Illy2005, Cussler2009, Welty2020}, as summarized in Table~\ref{tab:experiment}. Using these parameters we numerically solve the continuum model and compare it with the experimental result. The details of the numerical method can be found in Appendix~\ref{app:DEM}. The experimental parameters correspond to $\mathrm{Pe} \sim 10^5$ (using the length of the coffee bed and average velocity as characteristic length scale and velocity, respectively), a regime where advection clearly dominates. To the best of our knowledge, there is no estimated value for the mass transfer rate for dissolution. We therefore assume the mass transfer rate to be a free parameter that controls how fast the caffeine concentration decays in time. By varying the mass transfer rate and comparing the numerical solution with the experimental points we found a good quantitative agreement for $k_c=0.1~\textrm{m/s}$.

The solution of the continuum equation is plotted with a blue line in Fig.~\ref{fig:experiment}. The results obtained experimentally and from solving the continuum equation do overlap and the numerical result also reproduces the exponential decay of the concentration, except for the later extraction stages as discussed above. The concentration decays exponentially due to the time evolution of the concentration at the surface, Eq.~\eqref{eq:Cs_time}.

The inclusion of the liquid invasion, despite its simplicity, results in a monotonic decrease in the caffeine concentration with the extracted volume, as it is observed experimentally, and it avoids numerical artifacts due to the simultaneous dissolution of the grains~\cite{Mo2022}. The impact of the liquid invasion is further discussed in Appendix~\ref{app:invasion_importance}.

\subsection{Solute transport at the pore scale}

\begin{figure}
	\centering
	\includegraphics[width=\linewidth]{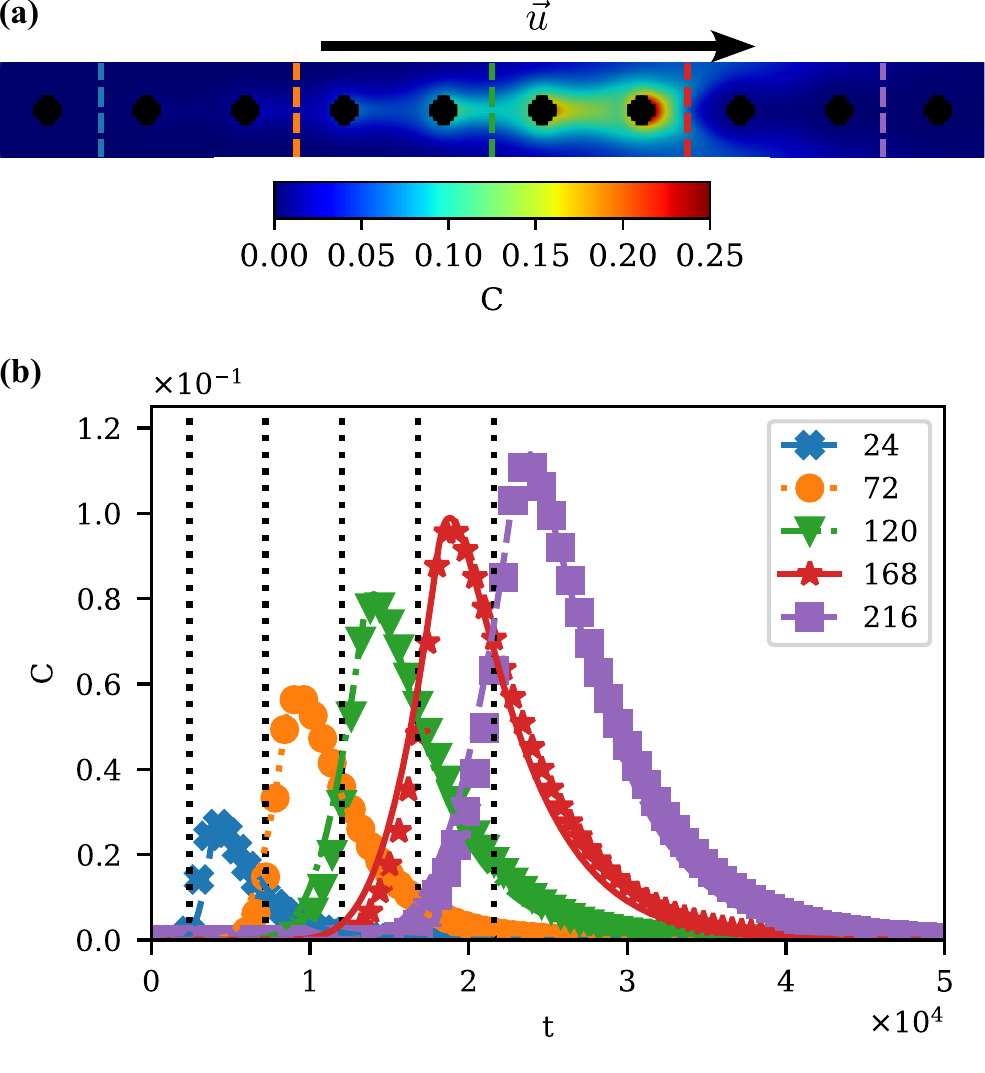}
	\caption{\textbf{(a)} Simulation domain with, in black, the grains where solute dissolves. The colours represent the solute concentration where blue (red) correspond to lower (higher) solute concentration, see colour scale. The fluid flows from left to right as indicated by the top arrow. The coloured vertical dashed lines correspond to the locations where the solute concentration was measured (see caption of (b)). \textbf{(b)} Solute concentration as a function of time in several measurement points (different colours). The labels correspond to the $ x $ position, in lattice units, of the measured point. The vertical black dotted lines correspond to the time $t^*$ at which the invading fluid reaches each measuring point for the first time.}
	\label{fig:interface}
\end{figure}

The results obtained with the continuum equation are in agreement with the experimental measurements for a regime of high Péclet number. To assess the validity of the continuum equation for a wider range of regimes we consider the dynamics at the level of the pore. The pore scale simulations also allow to test if the coarse grain approximations regarding dissolution, fluid velocity and hydrodynamic dispersion are valid. We start in the regime of low Péclet number that is representative of situations where the fluid flow is slower, such as gravity driven extractions, in aquifer contamination~\cite{Saripalli2001}. The pore scale simulations are conducted using the lattice Boltzmann method to solve both the fluid flow and the advection-diffusion equations, as described in Appendix~\ref{app:LBM}. The parameters used are listed in Table~\ref{tab:lbm}, all in lattice units (l.u.) a simple set of units scaled such that the timestep, lattice spacing, and fluid density, are $\Delta t = 1$, $\Delta x=1$, and $\rho=1$, respectively, and can be converted into real ones by comparing dimensionless numbers such as the Reynolds number. The diffusion coefficient is $1/3 \times 10^{-2}$ l.u., and the imposed inlet velocity is $10^{-2}$ l.u. The other boundaries are periodic, and we consider two-dimensional systems ($L_z=1$) thus, the grains are interpreted as infinite cylinders. Thus, advection dominates ($ \mathrm{Pe} \sim 10^2 $), but diffusion still plays an important role. The simulation domain is $ 240 \times 24 $ nodes and the circular grains have a radius of four nodes, as shown in Fig.~\ref{fig:interface}(a).

Figure~\ref{fig:interface}(b) shows the concentration as a function of time measured for different domain sites (colour), and different curves correspond to the same colour dashed line in Fig.~\ref{fig:interface}(a). At the inlet, the concentration is zero. As the fluid flows through the medium, the concentration increases as solute is dissolved and dragged by the fluid. Eventually the concentration reaches a maximum value when some areas of the grains have no solute to be dissolved anymore and the concentration starts to decrease as a result of a decrease in the dissolving area. The value of the maximum tends to $C_s(0,x)=C_i$, as most of the solute is concentrated at the fluid interface and dissolution, at the interface, ceases when $C=C_s$. The observed increase with the domain size is because the number of dissolving sources scales accordingly. Eventually the grains are fully depleted of solute and the concentration goes to zero.

The coloured lines in Fig.~\ref{fig:interface}(b) correspond to the numerical solution of the continuum equation. They are in agreement with the results of the pore scale. For this case, the hydrodynamic dispersion impacts the solute transport and it was used as the only fitting parameter, estimated using the Taylor-Airs dispersion~\cite{Taylor1953, Aris1956, Valdes-Parada2011}, Eq.~\eqref{eq:Deff_parallel}, with $\mathrm{Pe} \approx 10^2$. 

The vertical dashed lines in Fig.~\ref{fig:interface}(b) correspond to the time $t^*$ that takes for the invading fluid to reach the measuring point. The maximum concentration is always measured after $t^*$. With the increase of the length of the system, the maximum occurs later and later, when compared to $t^*$. Furthermore, the concentration starts to increase before $t^*$. This is because diffusion is driving the solute faster than the water invasion, which is expected with our simplified invasion implementation, since it only constrains the dissolution. Nevertheless, the approximation is valid for very large Péclet numbers.

\subsection{Limit of very low mass transfer rate}

In the limit of very low mass transfer rate, $k_c T_e / L \ll 1$, where $T_e$ is the extraction time, we can assume that $C_s$ does not vary in time, and the system is in a steady state. Thus, we can neglect the time dependence in Eq.~\eqref{eq:ade_adim} and, consequently, the water invasion, and assume that all grains start dissolving instantaneously.

Assuming constant velocity at the boundaries, unidimensional, and dissolution with infinite and homogeneous sources, Eq.~\eqref{eq:ade_adim} simplifies as,
\begin{equation}
	\mathrm{Pe} \frac{\partial C}{\partial x} = \frac{\partial^2 C}{\partial x^2} + \textrm{Sh} \xi (C_s - C) \textrm{ ,}
\end{equation}
where $x$ is the dimensionless length that varies between 0 (inlet) and 1 (outlet), and $\xi$ is the volume fraction of solute sources. The characteristic velocity and distance for the calculation of the Sherwood and Péclet numbers are the average velocity and length of the system, respectively. The boundary conditions are
\begin{equation}
	C(0) = 0 \textrm{ ,}
\end{equation}
and
\begin{equation}
	\frac{\partial C}{\partial x}(L) = 0 \textrm{ .}
\end{equation}
For the limit of pure diffusion or advection, \textit{i.e.} as $\mathrm{Pe}\ll1$ and $\mathrm{Pe}\gg1$, respectively, we obtain that the concentration at the outlet is
\begin{equation}
	C(L)=
	\begin{cases}
		C_s\left[1 - \frac{1}{\cosh(L\sqrt{\textrm{Sh}\xi})}\right] & \textrm{for }\mathrm{Pe} \ll 1 \\
		C_s\left(1 - e^{-L\textrm{Sh}\xi/\mathrm{Pe}}\right)\propto \mathrm{Pe}^{-1} & \textrm{for }\mathrm{Pe} \gg 1
	\end{cases}
	\textrm{ .}
	\label{eq:limits}
\end{equation}
See Appendix~\ref{app:full_solution} and Eq.~\ref{eq:full_solution} for further details on the solution.

\begin{figure}
	\centering
	\includegraphics[width=\linewidth]{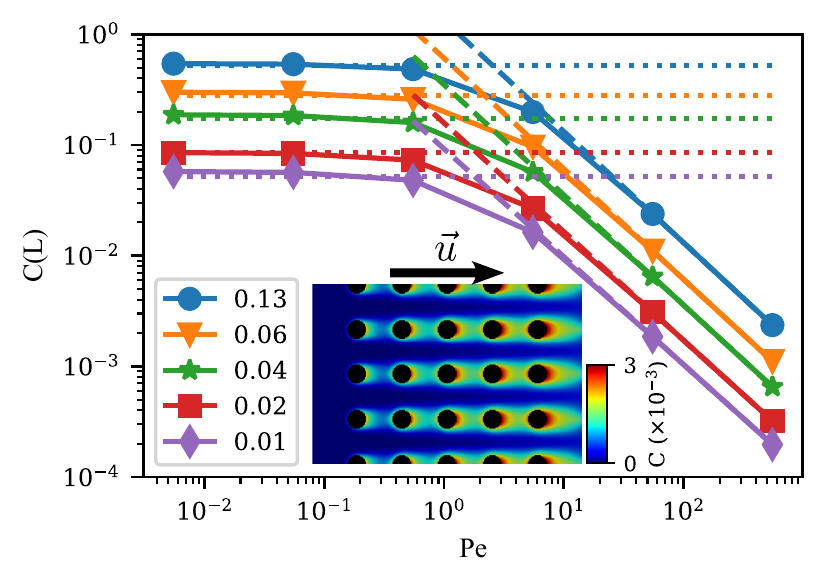}
	\caption{Solute concentration at the outlet of a grid of grains as a function of the Péclet number for different values of the density of grains (colours), $\xi=0.13$ corresponds to a domain with $12\times 13$ grains and $\xi=0.01$ to a domain with $2\times 3$ grains. The dots correspond to the pore scale simulations and the dashed and dotted lines to the limits of the analytical solution, Eq.~\eqref{eq:limits}, with the Sherwood number calculated from the hydrodynamic dispersion, Eq.~\eqref{eq:effective_dif}. The inset is the domain with circular obstacles (in black) surrounded by solute (colour). Red (blue) colour corresponds to higher (lower) concentration, see colour scale.}
	\label{fig:circle_sources}
\end{figure}

The fluid flow and solute transport at the pore scale is numerically solved for a domain containing a regular grid of solid disks, as represented in the inset of Fig.~\ref{fig:circle_sources}, and the results are compared with the analytical prediction, Eq.~\eqref{eq:limits}. The domain is $ 144 \times 216 $ nodes in size with evenly spaced circular grains of radius of four, an inlet velocity of $ 8 \times 10^{-2} $, and a mass transfer rate of $ 10^{-5} $. In Fig.~\ref{fig:circle_sources} it is plotted the solute concentration at the outlet of the domain ($L=216$) as a function of the Péclet number. For large Péclet numbers the concentration at the outlet decreases because it is being carried away by the fluid faster than the grains are injecting solute into the fluid. In this limit, the numerical results agree with the analytical prediction and the concentration decreases as Pe$^{-1}$. When the density of grains changes (different colours in Fig.~\ref{fig:circle_sources}) the concentration also decreases, simply due to the fact that a smaller fraction of sources injects less solute into the fluid. For low Péclet numbers the concentration is stable around a value that agrees with the analytical prediction (dotted line). This agreement is valid when we correct the diffusion coefficient to account for the geometry of the pores~\cite{McGreavy1992,Andrade1997}. Thus, we corrected the Sherwood number with an approximate expression for the diffusion in an array of circles~\cite{Dagdug2012}
\begin{equation}
	D^*=\frac{D}{\left(1-\frac{\pi}{4} \nu^2\right)\left[\nu \int_0^{\pi / 2} \frac{(\cos \varphi)^{1 / 3} d \varphi}{(1-\nu \cos \varphi)}+1-\nu\right]} \textrm{ ,}
	\label{eq:effective_dif}
\end{equation}
where $\nu=R/(l/2)$, $R$ the radius of the circles and $l$ is the distance between circle centers.

\section{Conclusions}
\label{sec:conclusion}

We proposed a continuum equation to study the transport of dissolved solute across porous media. We show that the proposed equation provides agrees with experiments, and with pore scale simulations of the fluid flow across arrays of circular grains. The proposed equation is valid for a large range of Péclet numbers, from the advection dominant regime $\mathrm{Pe}\gg1$ to the transition regime between advection and diffusion dominated transport $\mathrm{Pe}\sim 1$. We also show that the continuum equation agrees with experiments and pore scale simulations in a wide range of Sherwood numbers. The accurate results were obtained with a single fit parameter, namely, the mass transfer rate for the experimental result, and the hydrodynamic dispersion for the pore scale simulations. These two parameters can be measured experimentally, but we were unable to find such measurements for this specific case. For the case of very slow dissolution, all parameters are well defined and the results agree with the obtained analytical solution.

We show that the larger the volume of coffee extracted, the weaker it will be, as the maximum in caffeine concentration is at the very first drops of liquid. In spite of this, on the limit of low mass transfer rate, decreasing the fluid velocity increases the solute content of the extracted liquid, as we show that the solute concentration is inversely proportional to the Péclet number. For very low Péclet, the outlet concentration increases with the Sherwood number and surface area of the grains.

In this work we neglected the evolution of the solid matrix in the continuum model. This should impact the dissolution rate as it affects the surface area of the grains and the solute concentration at the surface. Furthermore, the changes to the fluid flow should also impact the subsequent solute transport. The good agreement with the experiential results suggests that the solid matrix evolution is not key in the extraction process, but that is not the case for extractions with novel materials such as swelling gels~\cite{Quesada-Perez2011}. Also, special care should be taken towards studying material with heterogeneous properties, as we assumed that the porosity and fluid velocity to be constant in the continuum model. In these cases, not only the Darcy fluid is space dependent, but the solid matrix evolution also varies in space. This will be address in future works.

More generally, our continuum equation can be adapted to model problems involving the sorption of solutes into the solid matrix. This allows to better understand and improve water filtration, increasing the availability of safe drinking water, specially in poorer communities~\cite{Mondal2019, Dalwadi2015, Dalwadi2016}. The case of filtering is similar to dissolution. The main differences are in the boundary conditions, namely, at the inlet the fluid contains solute, Eq.~\eqref{eq:inlet}, at the outlet it desirable clean fluid, Eq.~\eqref{eq:outlet}, and at the grain boundaries there is sorption of solute, Eq.~\eqref{eq:diss}, since the solute concentration on the liquid $C_l$ is larger than the concentration at the surface of the grains $C_s$. With this generalization, the appropriate physical behavior of filters, and in general absorbing media, should be captured. Also, the internal diffusion of caffeine inside the grains can be included in the model with an extra term on the evolution of the surface concentration, Eq.~\eqref{eq:Cs_time}, this should be enough to obtain accurate results for the caffeine concentration for long extractions.

\begin{acknowledgments}
	
We acknowledge financial support from the Portuguese Foundation for Science and Technology (FCT) under the contracts no. UIDB/00618/2020, UIDP/00618/2020, SFRH/BD/143955/2019, EXPL/FIS-MAC/0406/2021, and PTDC/FISMAC/5689/2020.

JSA acknowledges the Brazilian agencies CNPq, CAPES, and FUNCAP for financial support.

\end{acknowledgments}

\section*{Data Availability Statement}

The data that support the findings of this study are available from the corresponding author upon reasonable request.

\appendix

\section{Methods}
%\label{app:methods}

\subsection{Experimental realization}
\label{app:experiment}

\subsubsection{Chemicals and samples}

All reagents and solvents were analytical grade and used with no further purification. HPLC-grade methanol (\ch{MeOH}, $99.8\%$) and acetonitrile (\ch{ACN}, $99.8\%$) were purchased from Merck (Germany). Sodium chloride (\ch{NaCl}, $99.9\%$) and hydroxide (\ch{NaOH}, $98.0\%$) were obtained from AnalaR BDH Chemicals (UK). Hydrochloric acid ($37.0\%$) and sodium carbonate ($99.5\%$) were purchased from Riedel-de Haën (Germany). Ultra-pure water was obtained from Milli-Q water purification systems (USA). Coffee samples were obtained from Delta Q\texttrademark~(Portugal). The samples used were of the AromatiQ\texttrademark~variety in a capsule form and were used as obtained. All samples were filtered (Whatman No. 1 filters) and diluted by a 100 factor before analysis.

\subsubsection{Experimental set-up}

\textit{Coffee extraction procedure - }Using a Delta Q MilkQool Evolution\texttrademark~ coffee machine the extraction was carried using the long espresso program taking samples every $5$ mL. The espresso temperature was not controlled. In order to improve statistics, four extractions were carried with 10 samples of 5 mL each, obtained in sequence.

\textit{Soxhlet extraction procedure - }The coffee from one capsule was removed and weighted ($5.57943$ g), inserted into a cartridge and the extraction was carried out with $170$ mL of water for 24h. The SE apparatus was cleaned with $35$ mL of water giving a total of $205$ mL of extract.

\subsubsection{Instrumentation settings}

\begin{figure}
	\centering
	\includegraphics[width=\linewidth]{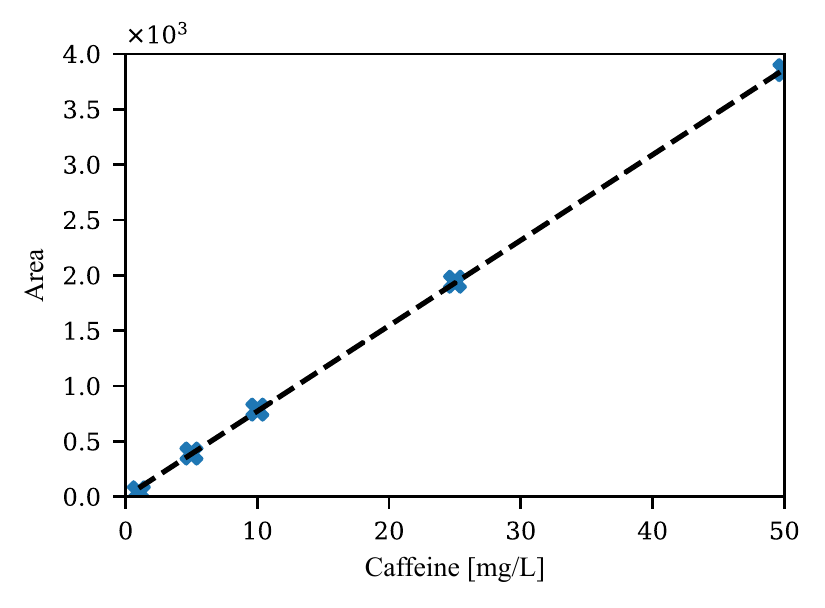}
	\caption{Integrated signal area, in arbitrary units, as function of the caffeine concentration. The dashed line corresponds to the trend line, Eq.~\ref{eq:hplc}.}
	\label{fig:hplc}
\end{figure}

HPLC-DAD analyses were carried out on an Agilent 1100 Series LC system (Agilent Technologies, Germany), constituted by the following modules: vacuum degasser (G1322A), quaternary pump (G1311A), autosampler (G1313A), thermostatted column compartment (G1316A) and diode array detector (G1315B). The data acquisition and instrumental control were performed by the software LC3D ChemStation (version Rev.A.10.02[1757], Agilent Technologies). 

Analyses were performed on a Kinetex C18 column, $150.0 \times 4.6$ mm, $2.6$ $\mu$m particle size (Phenomenex, Torrence, U.S.A.). The samples were analyzed using a mobile phase consisting of an isocratic mixture of $2.5\%$ acetic acid aqueous solution and \ch{ACN} ($85/15$, v/v). All solvents were previously filtered ($125$ mm of diameter, Cat. No. 1001 125, Whatman, U.K.) to remove possible interference particles. The detector was set at 278 nm and the column temperature at $20~\textrm{ºC}$. The injection volume was $10$ $\mu$L with a draw speed of $200$ $\mu$L min$^{-1}$ and the flow rate was set at $0.6$ mL min$^{-1}$.

The HPLC calibration was performed with six caffeine standard solutions having concentrations ranging from $1.0$ to $50.0$ mg$\cdot$L$^{-1}$. From the data obtained, a linear dynamic response was achieved for caffeine with a correlation coefficient of $0.9999$, Fig.~\ref{fig:hplc}. The integrated signal area ($I$) of the peak was used to obtain a calibration curve
\begin{equation}
	I = 77.221[\textrm{L}\cdot\textrm{mg}^{-1}]\cdot[\textrm{caffeine}] \textrm{ ,}
	\label{eq:hplc}
\end{equation}
where $[\textrm{caffeine}]$ is the caffeine concentration in mg$\cdot$L$^{-1}$.

\subsection{Discrete Element Method}
\label{app:DEM}

The continuum equation was solved using one dimensional central finite differences (second order) and with an Euler scheme. To compare with the experimental realization, the grid spacing was $ 10^{-4}~\textrm{m}$ and the timestep $5\times 10^{-3}~\textrm{s}$. For the comparison with the pore scale simulations the grid space and the timestep were both set to two.

\begin{table}
	\centering
	\caption{List of parameters measured experimentally, in previous works (references in the table), and considered to solve Eq.~\eqref{eq:ade_hom}.}
	\begin{tabular}{lcr}
		\hline\hline
		Parameter                     &          Symbol           &                                 Value \\ \hline
		Cake height                   &            $L$            &          $10^{-2}$ m~~\cite{Illy2005} \\
		Fluid velocity                & $\lVert \mathbf{u}_m \rVert$ &                         $10^{-2}$ m/s \\
		Diffusion coefficient         &            $D$            & $10^{-9}$ m$^2$/s~~\cite{Cussler2009} \\
		Mass transfer rate            &           $k_c$           &       $10^{-1}$ m/s~~\cite{Welty2020} \\
		Total caffeine mass           &           $m_T$           &                                 70 mg \\
		Initial surface concentration &           $C_i$           &            $1.4 \times 10^5$ mg/m$^3$ \\
		Outlet radius                 &           $R_e$           &                   $4\times 10^{-3}$ m \\
		Extraction time               &           $T_e$           &                                $50$ s \\ \hline\hline
	\end{tabular}
	\label{tab:experiment}
\end{table}

\subsection{Pore scale simulation}
\label{app:LBM}

\subsubsection{Lattice Boltzmann method}

The lattice Boltzmann method~\cite{Kruger2017, Succi2018} allows fluid simulations in complex geometries, as it is the case of porous media, and can also be used to solve the advection-diffusion equation. We consider incompressible fluids flows
\begin{equation}
	\nabla \cdot \mathbf{u} = 0 \textrm{ ,}
\end{equation}
and the fluid flow is governed by the Navier-Stokes equation
\begin{equation}
	\dot{\mathbf{u}}+(\mathbf{u} \cdot \nabla) \mathbf{u} = \nu \nabla^{2} \mathbf{u}-\frac{1}{\rho} \nabla p \textrm{ ,}
	\label{eq:NSeq}
\end{equation}
where $\nu$ is the kinematic viscosity, $\rho$ is the fluid density and $p$ is the fluid pressure. In the macroscopic limit, the lattice Boltzmann method recovers the Navier-Stokes equation, Eq.~\eqref{eq:NSeq}, by discretizing the Boltzmann equation and solving it numerically on a regular grid. We use the D3Q19 discretization, where space is discretized into a regular cubic grid in 3D, of linear lengths $L_x$, $L_y$ and $L_z$, and velocity is discretized into a set of 19 velocity vectors. Despite using a three-dimensional scheme, all simulations have $L_z=1$ thus, it is effectively two dimensional. The discretized Boltzmann equation is
\begin{equation}
	\frac{f_i(\mathbf{x}+\mathbf{c}_i \Delta t, t + \Delta t) - f_i(\mathbf{x}, t)}{\Delta t} = \left(\frac{\partial f_i}{\partial t}\right)_\textrm{coll}\textrm{ ,}
\end{equation}
where $f_i(\mathbf{x}, t)$ is the distribution function in the direction $\mathbf{c}_i$ at the node with position $\mathbf{x}$ and time $t$, $\Delta t$ is the time step, $\mathbf{c}_i$ are the discretized velocities, and the right-hand side accounts for the collision between particles. For the collision term, we consider the two-relaxation-time operator (TRT). This method splits the distribution into a symmetric $f_i^+$ and an antisymmetric part $f_i^-$
\begin{equation}
	f_i^+ = \frac{f_i + f_{\bar{i}}}{2} \quad \textrm{and} \quad f_i^- = \frac{f_i - f_{\bar{i}}}{2} \textrm{ ,}
\end{equation}
where $f_{\bar{i}}$ correspond to the distribution going to the opposite direction of $f_i$. The BGK collision term with two relaxation times $\tau^+$ and $\tau^-$ is employed to relax towards equilibrium the symmetric and antisymmetric distribution, respectively. The two relaxation times are related via
\begin{equation}
	\Lambda = \left(\frac{1}{\tau^+\Delta t} - \frac{1}{2}\right)\left(\frac{1}{\tau^-\Delta t} - \frac{1}{2}\right) \textrm{ ,}
	\label{eq:TRT_Lambda}
\end{equation}
where $\Lambda$ is a parameter that allows to control the accuracy and stability of the method~\cite{Ginzburg2005a}. Fixing $\Lambda = 3/16$ results in the boundary wall location implemented via bounce-back for the Poiseuille flow to be exactly in the middle between solid and fluid nodes~\cite{Ginzburg2008, Kruger2017}. The first relaxation time sets the viscosity $\nu$ of the fluid
\begin{equation}
	\nu=c_s^2\left(\tau^+ - \frac{\Delta t}{2}\right) \textrm{ .}
\end{equation}
We set $\tau^+=0.8$, which corresponds to $\nu=0.1$ l.u.. This parameter, together with the velocity range considered, results in a laminar flow. The second relaxation time $\tau^-$ is determined with Eq.~\eqref{eq:TRT_Lambda}. With the collision operator the fluid distributions evolve towards the equilibrium distribution $f^\textrm{eq}_i$
\begin{equation}
	f^\textrm{eq}_i = w_i \rho \left[1+\frac{\mathbf{c}_i\cdot\mathbf{u}}{c^2_s} - \frac{\mathbf{u} \cdot \mathbf{u}}{2c^2_s} + \frac{(\mathbf{c}_i\cdot\mathbf{u})^2}{2c^4_s}\right] \textrm{ ,}
\end{equation}
where $w_i$ are weights corresponding to the discretized velocities ($w(\lVert \mathbf{c}_i \rVert=0)=1/3$, $w(\lVert \mathbf{c}_i \rVert=1)=1/18$, and $w(\lVert \mathbf{c}_i \rVert=\sqrt{2})=1/36$), $\rho$ is the fluid density, and $c_s$ is the speed of sound, which for the D3Q19 is $1/\sqrt{3}$. The macroscopic variables of the fluid  (density $\rho$ and velocity $\mathbf{u}$) are determined from the distribution functions with $\rho = \sum_i f_i$ and $\rho \mathbf{u} = \sum_i f_i \mathbf{c}_i$.

On the solid-liquid interface, no slip boundaries are imposed with an interpolated bounce-back method, proposed by Mei \textit{et al.}~\cite{Mei2002}. The inlet and outlet velocities are fixed by imposing the corresponding distributions, as proposed by Kutay \textit{et al.}~\cite{Kutay2006}.

The solid matrix is defined by nodes that cover the solid region. Each node has a mixture of fluid and solid measured by the solid fraction $m(\mathbf{x})$~\cite{Matias2021, Jager2017}. Depending on the solid fraction of each node, there are three types of nodes: nodes that contain only fluid, nodes that contain only solid, and interface nodes that contain both solid and fluid. This translates into a scalar field
\begin{equation}
	m(\mathbf{x})=
	\begin{cases}
		0     & \quad \text{fluid node,}\\
		]0,1] & \quad \text{interface node,}\\
		1     & \quad \text{solid node.}
	\end{cases}
\end{equation}
If $m(\mathbf{x})=1$ for a node, then that node is considered interface or solid depending if it has a fluid neighbor or not, respectively.

\begin{table}
	\centering
	\caption{List of parameters, in lattice units, considered in the lattice Boltzmann method and to solve Eq.~\eqref{eq:ade_hom}.}
	\begin{tabular}{lcr}
		\hline\hline
		Parameter                     &            Symbol            &                Value \\ \hline
		Cake height                   &             $L$              &                  240 \\
		Cake width                    &            $L_y$             &                   24 \\
		Fluid velocity                & $\lVert \mathbf{u}_m \rVert$ &            $10^{-2}$ \\
		Diffusion coefficient         &             $D$              & $1/3 \times 10^{-2}$ \\
		Diffusion coefficient         &            $D^*$             & $2/3 \times 10^{-1}$ \\
		Mass transfer rate            &            $k_c$             &            $10^{-3}$ \\
		Initial surface concentration &            $C_i$             &                  $1$ \\
		Extraction time               &            $T_e$             &      $5 \times 10^4$ \\ \hline\hline
	\end{tabular}
	\label{tab:lbm}
\end{table}

\subsubsection{Lattice Boltzmann method for advection-diffusion problems}
\label{app:theory_ade}

The advection-diffusion equation, Eq.~\eqref{eq:ade}, can also be numerically integrated with the lattice Boltzmann method~\cite{Kruger2017}. We considered two coupled lattice Boltzmann algorithms: one solves the fluid flow, and the other takes the velocity field from the first and solves the solute transport. For the collision term, we consider the BGK collision operator, that considers only one relaxation time $\tau_g$. When using the lattice Boltzmann method for advection-diffusion problems the relaxation time sets the diffusion coefficient
\begin{equation}
	D = c_s^2\left(\tau_g - \frac{\Delta t}{2}\right) \textrm{ .}
	\label{eq:lbm_dm}
\end{equation}
It was considered a low relaxation time $\tau_g = 0.51$ which corresponds to a diffusion of $1/3 \times 10^{-2}$, as indicated in Table~\ref{tab:lbm}. The coupling between the two LBM solvers occurs on the calculation of the solute equilibrium distribution
\begin{equation}
	g_i^\textrm{eq} = w_i C \left[1+\frac{\mathbf{c}_i\cdot\mathbf{u}}{c^2_s} - \frac{\mathbf{u} \cdot \mathbf{u} }{2c^2_s} + \frac{(\mathbf{c}_i\cdot\mathbf{u})^2}{2c^4_s}\right] \textrm{ ,}
\end{equation}
where $C$ is the solute concentration of the node, and $\mathbf{u}$ is the fluid velocity at the node, obtained with the fluid solver. We assume that the solute is passively transported, thus fluid viscosity and the boundary locations remain constant~\cite{Nayar2016, Moore2017, Quirk1955}.

\begin{figure}
	\centering
	\includegraphics[width=.9\linewidth]{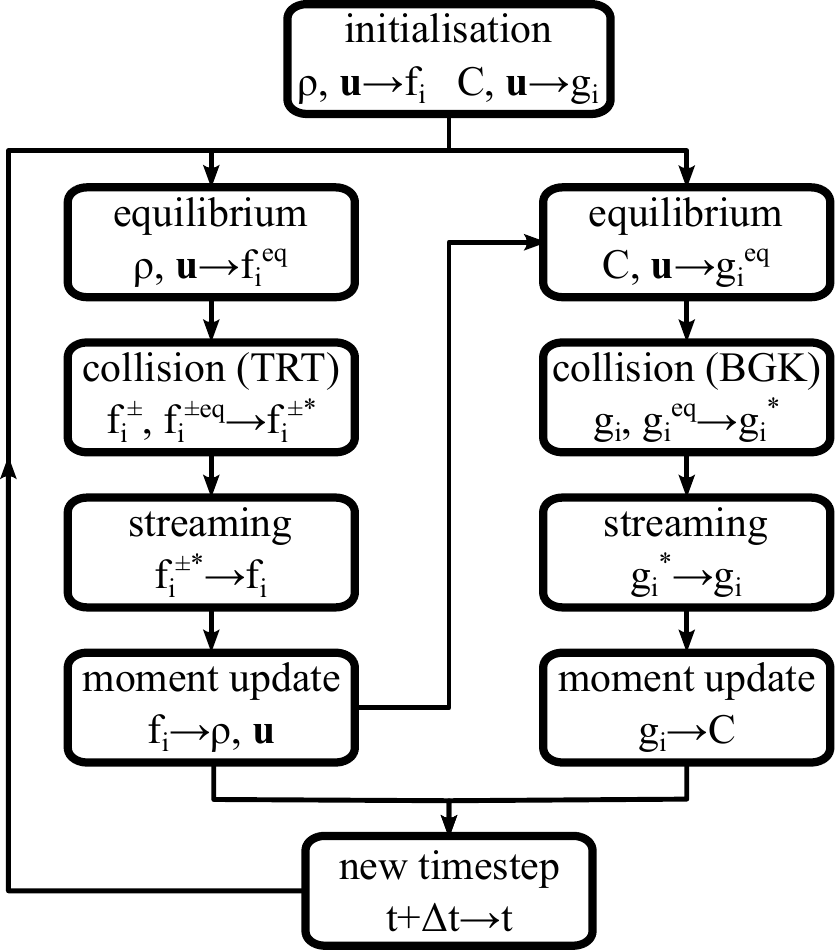}
	\caption{Workflow representation of the coupled lattice Boltzmann algorithms.}
	\label{fig:workflow}
\end{figure}

We use the bounce-back boundary condition for inert obstacles \textit{i.e.} no flux of solute through their surface. At the inlet, we impose a concentration equal to zero with the anti-bounce-back boundary condition~\cite{Kruger2017}. At the outlet, an open boundary is imposed by setting the gradient of concentration normal to the boundary to zero~\cite{Kruger2017}. For the dissolution of solute across the solid-liquid interface, we include a source term on the interface nodes. This approach is simpler than implementing directly the spatial derivative, as in Refs.~\cite{Ju2020, Ginzburg2005b, Yoshino2003}, and it allows for a more accurate measurement of the amount of mass that dissolves. The simplest way is to add the source term $S$, Eq.~\eqref{eq:diss}, during the collision step
\begin{equation}
	g_i(\mathbf{x}+\mathbf{c}_i \Delta t, t + \Delta t) = g_i(\mathbf{x}, t) + \left(\frac{\partial g_i}{\partial t}\right)_\textrm{coll} + \left( 1 - \frac{1}{2\tau_g} \right) w_i S \textrm{ ,}
	\label{eq:coll_with_source}
\end{equation}
and to correct the calculation of the macroscopic concentration
\begin{equation}
	C = \sum_i g_i + \frac{S \Delta t}{2} \textrm{ .}
\end{equation}
Notice that, this implementation is similar to the inclusion of a force term in the traditional LBM, for the velocity field. To calculate the source term, Eq.~\eqref{eq:source_term}, we consider that $A$, $V_l$ and $V_s$ are the surface area, liquid and solid volume of each interface node that is dissolving. The worlfow for the two algortimts is represented in Fig.~\ref{fig:workflow}.

\subsubsection{Surface area}

To estimate the surface area of the solid-fluid interface at each node, we assume that the interface is composed solely of planes whose orientation, and consequently their area, only depends on the normal to the surface of an interface. This approach is a simplified version of the one introduced in Ref.~\cite{Kashani2022}, but produces accurate results nonetheless. A detailed description of the algorithm can be found in Ref.~\cite{Matias2021}. The summary is as follows, the normal vector to the surface is the normalized weighted average of the neighboring mass volumes $V_s$
\begin{equation}
	\hat{n}(\mathbf{x})=\frac{\sum_i V_s(\mathbf{x}+\mathbf{c}_i)\mathbf{c}_i w_i}{\lVert\sum_i V_s(\mathbf{x}+\mathbf{c}_i)\mathbf{c}_i w_i \rVert} \text{ ,}
	\label{eq:normal}
\end{equation}
where the sum is over all velocities $\mathbf{c}_i$. With the vector normal to the surface, the surface area of the interface node is
\begin{equation}
	A = \frac{\sqrt{1+n_2^2}}{(1+n_2)} \frac{\sqrt{1+n_3^2}}{(1+n_3)} \text{ .}
	\label{eq:surface_area}
\end{equation}
where $n_i$ are the components of the normal vector scaled by the smallest non-zero component such that $n_1=1$ and $n_1\leq n_2 \leq n_3$.

\subsubsection{Numerical validation of the LBM}

\begin{figure}
	\centering
	\includegraphics[width=\linewidth]{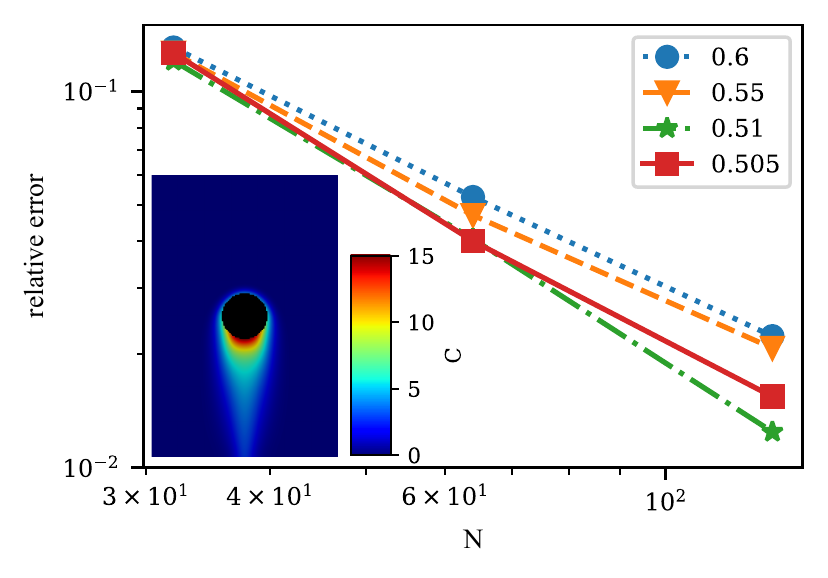}
	\caption{Relative error of the injected mass as a function of the domain length $N$ for different relaxation times $\tau_g$ (markers and colours), see Eq.~\eqref{eq:lbm_dm}. The relative error is calculated using the theoretical mass rate, Eq.~\eqref{eq:mdot_teo}, and the rate at which the mass exists the domain $\dot{m}_n$, Eq.~\eqref{eq:mdot_num}. The inset show the spatial distribution of the solute on a domain with $N=128$ with the fluid flowing from top to bottom. The colours represent the solute concentration, blue (red) for low (high) concentration, see colour scale.}
	\label{fig:mass_error}
\end{figure}

\begin{figure}
	\centering
	\includegraphics[width=\linewidth]{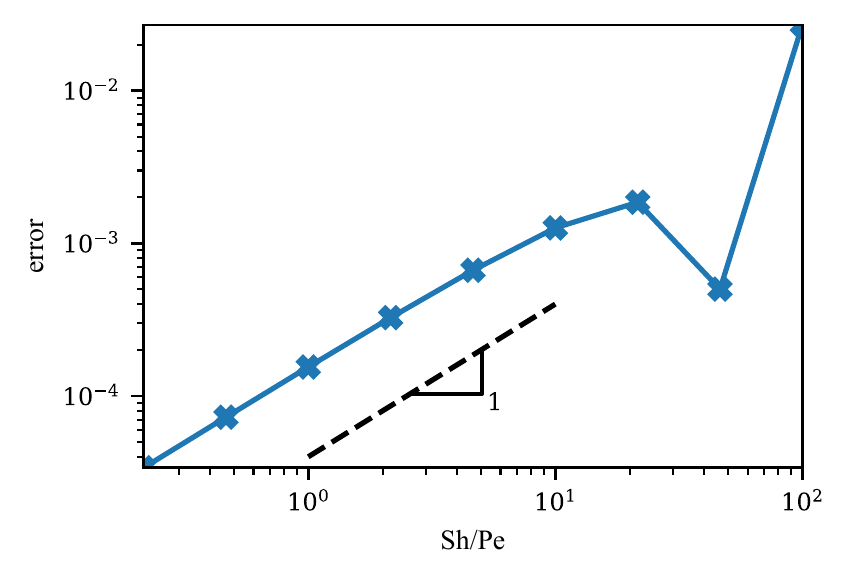}
	\caption{Plot of the concentration error, Eq.~\eqref{eq:C_error}, as function of the ratio between Sherwood and Péclet number. The Péclet number is fixed at $10^5$. The effect of the inclusion of the fluid invasion is more relevant the larger the Sherwood number is.}
	\label{fig:error_interface}
\end{figure}

The inclusion of the source term, as detailed in Sec.~\ref{sec:theory_diss} and Appendix~\ref{app:theory_ade} was validated by simulating the solute output on a domain, with dimensions $N \times 1.5N$, containing a circle of radius $N/8$ in the center, as represented on the inset of Fig.~\ref{fig:mass_error}. The fluid inlet velocity is $ 5\times 10^{-2} $ and the mass transfer rate $ 5\times10^{-3} $. Since this test aims at verifying the correct implementation of the source term, we consider that the dissolution does not depend on the liquid concentration. This is valid when the solute concentration on the fluid is much smaller than the saturation concentration. Thus
\begin{equation}
	S = k_c C_s \frac{A}{v_l} \textrm{ ,}
\end{equation}
where $C_s = 1$ is fixed. Given that the dissolution does not depend on time we can evaluate the solute spread at steady state. Since we have mass conservation, the mass injected into the fluid must equal the mass that exits the domain. The total mass injected on the fluid, per unit time is
\begin{equation}
	\dot{m}_t=k_c C_s A \textrm{ ,}
	\label{eq:mdot_teo}
\end{equation}
where $A$ is the area of the circle, and the mass that exits the domain is given by
\begin{equation}
	\dot{m}_n = \int C \mathbf{u} \cdot d\mathbf{A} \textrm{ ,}
	\label{eq:mdot_num}
\end{equation}
where the integral is over the outlet nodes. By comparing the two values we can evaluate the accuracy of the boundary conditions. Figure~\ref{fig:mass_error} is the plot of the relative error as a function of the domain size for several relaxation times, \textit{i.e.} diffusion coefficients. As expected, increasing the domain size decreases the error, since the discretization of the circle is finer. Changes in the relaxation time have little influence on the error, and with the few points we have, there is no clear trend. Notice, however, that with relaxation times bellow $0.505$ the numerical stability start to be compromised.

\section{The liquid invasion}
\label{app:invasion_importance}

To evaluate the importance of modeling the fluid invasion, Eq.~\eqref{eq:ade_hom}, we simulated the solute transport with and without the inclusion of the fluid invasion (Heaviside function). We solved, using finite differences, Eq.~\eqref{eq:ade_hom} and \eqref{eq:Cs_time} for a one-dimensional domain with length $L = 10^{-2}$, fluid velocity $\lVert\mathbf{u}_m\rVert = 10^{-2}$, diffusion coefficient $D=10^{-9}$, $m_T=1$ in a grid with a spacing of $10^{-4}$ and a timestep of $10^{-4} T_e$, where $T_e$ is the extraction time. The extraction time is $T_e=L/\lVert\mathbf{u}_m\rVert - \log(10^{-2})/k_c$, this ensures we always simulate a full extraction. The choice of parameters results in $\mathrm{Pe}=10^5$, Eq.~\eqref{eq:peclet}, and we varied the mass transfer rate $k_c\in[10^{-3}, 10^0]$ to change the Sherwood number, Eq.~\eqref{eq:sherwood}.

To evaluate the error of the numerical result we compared the solute concentration at the outlet with the fluid invasion $C$ and without the invasion $C'$. Thus the relative error is
\begin{equation}
	\textrm{error} = \int_0^{T_e} \frac{C - C'}{C} dt \textrm{ .}
	\label{eq:C_error}
\end{equation}
In figure~\ref{fig:error_interface} we plot the error as function of the Sherwood number. The error increases with the mass transfer rate and stabilizes for $\textrm{Sh}/\mathrm{Pe}=10^2$, which corresponds to $k_c=1$, meaning the solute dissolves instantly (notice that $C_s(t=0)=1$). The error increases linearly. This is expected, as the modeling of the invading fluid should be more important for situations where the mass transfer rate is fast, when compared to the invasion velocity, since most of the solute will concentrate at the fluid front. As we saw in Sec.~\ref{sec:coffee}, in the case of espresso extraction ($\textrm{Sh}/\mathrm{Pe}\sim10$), most of the solute is at the invading front thus the modeling of the invading fluid is important in the accuracy of the results, in particular for the initial times.

\section{Analytical solution of point sources}
\label{app:full_solution}

\begin{figure*}
	\centering
	\includegraphics[width=\linewidth]{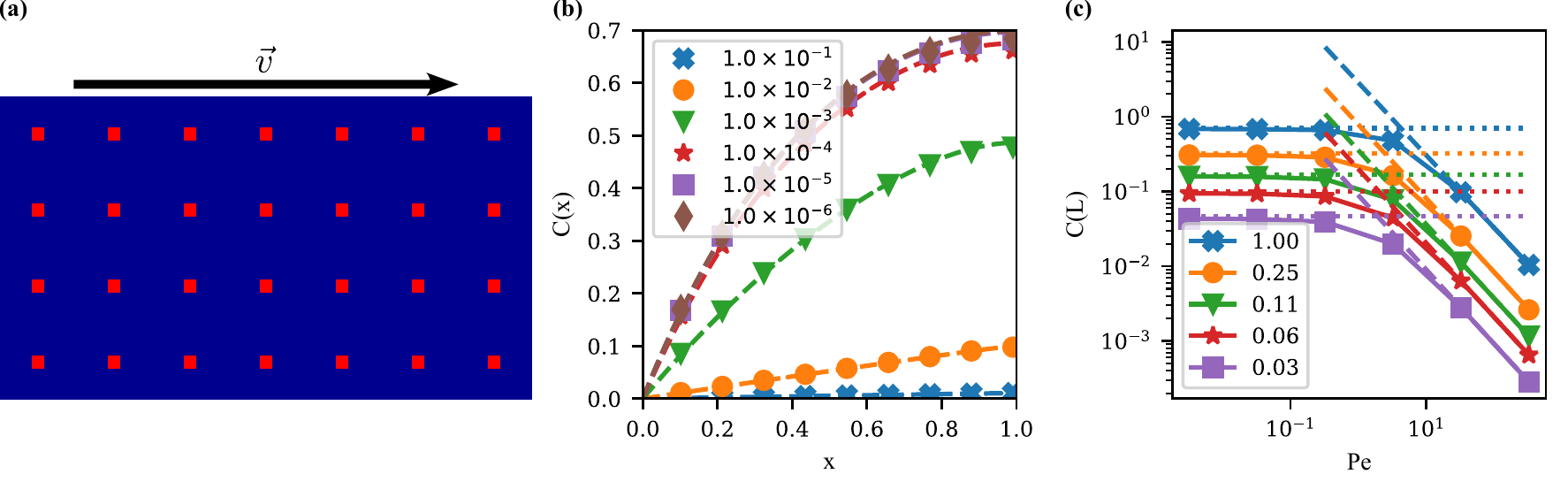}
	\caption{\textbf{(a)} Domain with point sources (red squares) spaced every 6 nodes. \textbf{(b)} Solute concentration as a function of position, the dashed line corresponds to the analytical solution of the advection-diffusion equation, Eq.~\eqref{eq:ade}, and the dots were obtained with the LBM. Different colours correspond to different fluid velocities. \textbf{(c)} The dots are the solute concentration at $x=108$ obtained with LBM, dashed (dotted) lines correspond to the limit $\mathrm{Pe}\ll 1$ ($\mathrm{Pe}\gg 1$) of the analytical solution for, Eq.~\eqref{eq:limits}. Each colour correspond to a different density of point sources $\xi$, where $\xi=1$ corresponds to a domain filled with sources and $\xi=0.03$ to a domain with sources spaced every 6 nodes, as represented in \textbf{(a)}.}
	\label{fig:point_sources}
\end{figure*}

For very low mass transfer rate and point-like sources the governing equation is reduced to
\begin{equation}
	\mathrm{Pe} \frac{\partial C}{\partial x} = \frac{\partial^2 C}{\partial x^2} + \textrm{Sh} \xi (C_s - C) \textrm{ .}
\end{equation}
This can be solved analytically obtaining
\begin{widetext}
	\begin{equation}
		C(x) = C_s \frac{\Delta \left(1 +e^{\Delta} -e^{S_+ x} -e^{S_- x + \Delta}\right)-\mathrm{Pe} \left(1 -e^{\Delta} -e^{S_+ x} +e^{S_- x + \Delta}\right)}{\Delta \left(1+e^{\Delta}\right)-\mathrm{Pe} \left(1-e^{\Delta}\right)}
		\textrm{ ,}
		\label{eq:full_solution}
	\end{equation}
\end{widetext}
where $\Delta = \sqrt{\mathrm{Pe}^2+4\textrm{Sh}\xi}$ and $S_\pm=(\mathrm{Pe}\pm\Delta)/2$.

We compare this analytical solution with pore scale simulations of a domain with point sources, as illustrated in Fig.~\ref{fig:point_sources}(a). The point sources are evenly spaced in a domain of $ 72 \times 108 $ nodes that dissolve with a mass transfer rate of $ 10^{-5} $. In Fig.~\ref{fig:point_sources}(b) we observe a good agreement between the analytical solution (dashed lines) and the pore scale simulations (points). The solute concentration across the system depends on the competition between the fluid velocity (different colours), that carries the solute, and the mass transfer rate, that injects solute. Without fluid velocity the solute concentration converges exponentially fast towards the surface concentration $C_s$. On the opposite case, as the velocity increases to infinity the solute concentration should go to zero, as it is moved out of the domain increasingly faster. In Fig.~\ref{fig:point_sources}(c) it is plotted the solute concentration at the outlet of the domain ($L=108$) as a function of the Péclet number and compares with the analytical solution, Eq.~\eqref{eq:limits}. Like for the case of a domain with circles, the pore scale simulations agree with the analytical solution. For this case there is no need to correct the diffusion coefficient of the Sherwood number.

\nocite{*}
\bibliography{ref}% Produces the bibliography via BibTeX.

\end{document}